# Two-Dimensional Material Based Photodetector for Near Infrared Detection

Farhad Larki[1,2,3], Arash Dehzangi[4], Alam Abedini[1,5], Muhamad Ramdzan Buyong[3], Hossein Tavakol[5], Parviz Kameli[2], Hadi Salamati[2]

[1]Laye Rooyan Part, Isfahan Science and Technology Town, Isfahan, Iran. [2]Department of Physics, Isfahan University of Technology, Isfahan 84156, Iran. [3]Institute of Microengineering and Nanoelectronics (IMEN), Universiti Kebangsaan Malaysia, Bangi, Selangor, Malaysia. [4]Department of Electrical and Computer Engineering, University of Texas at Dallas, Richardson, Texas, USA. [5]Department of Chemistry, Isfahan University of Technology, Isfahan, Iran

**Abstract:** Implementing new materials as alternative to silicon for application in photonic devices has been the center of attention in the scientific community. Two-Dimensional (2D) materials have shown a great capacity to be next alternative to remedy this impediment. Graphene due to its unique properties such as high mobility and optical transparency, in addition to flexibility, robustness and environmental stability is amongst 2D materials to shine in photonics and optoelectronics. Several graphene-based photodetectors with ability to integrate with various energy thermoelectric, electromagnetic and piezoelectric devices is reported. However, pristine graphene is not suitable to detect reasonable signals in infrared region, due to its bandgap limitation. In this work, a graphene based near infrared detection is demonstrated using graphene/metal intercalated graphene photodetector. The intercalated graphene was grown using chemical vapor deposition (CVD) on Cu substrate and wet transfer of the layers to the Si/SiO2 substrate. A tapered aluminum microelectrode has been used for the electrical contacts to improve the detection of photogenerated carriers during the illumination. The infrared detection was demonstrated, responsivity and quantum efficiency was tested at room temperature and the mechanism of photogeneration was explained.

## INTRODUCTION

Photodetectors (PDs), which convert light into electrical signals, play a crucial role in various fields such as data transmission [1], night vision imaging [2-4], wearable devices [5], and automotive radar system [6]. The main challenge for Si PDs which are the dominate technology in the past few years, is due to its band gap width of 1.1 eV and the limited coverage of detection to cover broad band application beyond visible light region. When it comes to near infrared (NIR) and short–wavelength infrared (SWIR) several material systems have been used for detection; each of those material systems has its own advantages and disadvantages. II-IV material system, e.g. Mercury–cadmium–telluride (HgCdTe) is a strong material system with ability to realize photodetectors that cover the entire infrared spectral range, [7-10] but certain factors, such as complex material growth and device fabrication processes [11, 12] are limiting complete domination of this material system over the IR detection technology. When it comes to III-V materials, high performance $In_xGa_{1-x}As$ compounds photodetectors when lattice–matched to InP substrate are great candidate for NIR and SWIR photodetection, but their performance sometimes are drastically affected by mismatch–induced defects for certain wavelengths [13, 14]. A strained layer superlattice III-V based material is also another strong candidates for infrared photodetection [15-18], this material is under development to deliver high performance for NIR and SWIR detection [3, 4].

Various two dimensional (2D) materials such as transition metal disulfide (TMD) [19], metal nitrides/carbonitrides (MXenes) [20], hexagonal boron nitride (h-BN) [20], perovskite [21] and graphene [22] has been implemented in photodetectors to improve the photodetection and imaging in the broad range of electromagnetic spectrum and infrared detection [23, 24]. Graphene and graphene-based materials exhibit exceptional optical and electrical properties with great promise for novel applications in photo detection. Graphene provides an opportunity for low cost, large area, broad band photodetectors because of its high-speed operation (∼1.5 THz), very low carrier effective mass and high mobility ($2 \times 10^5$ cm$^2$ V$^{-1}$ S$^{-1}$) [25-27]. However, a single layer graphene (SLG) absorbs only 2.3% of light, and because of its zero-bandgap nature, the dark current is very high which is a serious limitation for practical applications.

Alternation structures and mechanism, such as forming graphene heterojunctions with other 2D materials [28, 29], introducing lateral confinement [30], applying an electric field perpendicular to a heterojunction [31] and chemical doping [32] have been proposed for band gap opening in graphene. Among various graphene doping methods, intercalation doping of graphene with ferric chloride ($FeCl_3$) has been theoretically and experimentally confirmed to be an effective method to reduce the sheet resistance of few-layer graphene, tune the fermi surface of graphene while maintain its transparency [33-36]. $FeCl_3$ intercalated few-layer graphene (IFLG) also is highly stable in ambient conditions as well as high humidity and temperatures environment and can be used to define photo-active junctions. Here we took the advantage of magnetic field assisted pulsed laser deposition (MFPLD) technique for growth of high quality FLG and intercalated CVD grown graphene with $FeCl_3$ to fabricate a graphene-based device to operate in near infrared (NIR) region of electromagnetic spectrum. The details of fabrication will be addressed and the operation of photodetector analyzed and discussed based on the mechanism of photodetection and charge transfer in the active region of illuminated device.

## MATERIAL AND METHODS

### Device Fabrication

The pristine graphene growth performed through magnetic field assisted pulsed laser deposition (MFPLD) technique. The schematic of the growth chamber is presented in Figure 1a. The details of the growth and the advantage of this technique have been mentioned in ref [37].

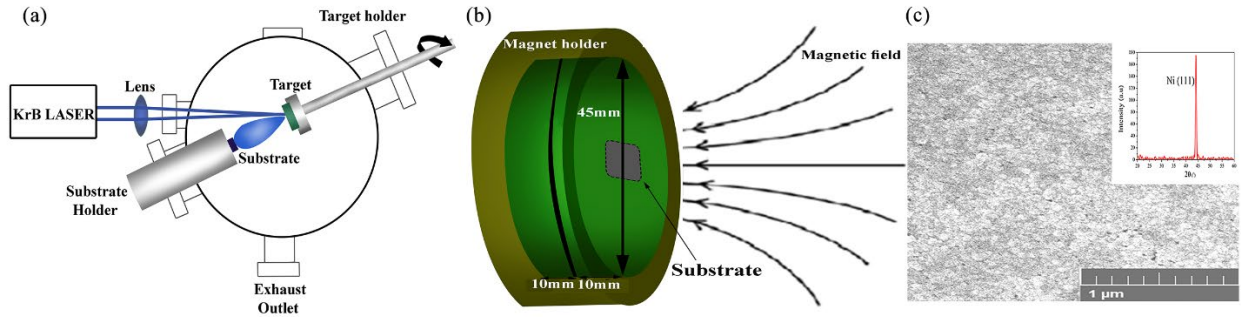

Figure 1. Schematic of the (a) PLD system. (b) permanent magnets configuration used for Nickle and graphene growth and (c) top view of Ni layer grown with PLD under the application of magnetic field.

A Lambda Physik pulsed KrF excimer laser ($\lambda$ = 248 nm) was used throughout. The process starts with preparing a $Si/SiO_2$ (~300 nm)/Ni (~300 nm) substrate. Since the quality of the Ni surface is one of the most important factors which determines the graphene quality, the Ni layer grown using PLD through application of a stationary magnetic field with 20.000 shot of laser (10 Hz, 230 mJ) in background pressure of $2 \times 10^{-6}$ Torr at room temperature under the influence of a 180 mT NdFeB (N35) permanent magnet (Figure 1b). The top view of Ni is shown in Figure 1c. In order to grow the graphene, a rotating graphite target ablated by the same laser with 500 shot (100 mJ at 4Hz) under the effect of a NdFeB permanent magnets with surface magnetic flux densities of about 350 mT.

### CVD graphene

Chemical vapour deposition (CVD) technique is a promising and well-developed technique for the large area graphene growth. Through careful selection of substrate, SLG and FLG can be grown [38]. Before the CVD process, the Cu substrates were prepared for deposition through ultrasonicating the samples in acetone, ethanol and isopropanol. Two main parameters of deposition, which are temperature of deposition, and the time of gas flow have been studied and optimized. The flow times of 180 s and deposition temperature of 1000 °C have been chosen according to the optimization process.

### $FeCl_3$ Intercalation

The intercalation of graphene is performed by two zone evaporation technique [34, 39]. Graphene and the intercalants placed in two regions of a designed ampoules and pumped to $1 \times 10^2$ Torr vacuum, then sealed in order to prevent oxidation of the samples.
 Graphene zone and anhydrous $FeCl_3$ powder zone kept at a constant temperature of 360°C and 310°C respectively, for 12 hours and then cooled down to room temperature naturally.

**Tapered aluminium microelectrode (TAM) fabrication**

The fabrication of tapered aluminium microelectrode (TAM) is based on CMOS processing technique. The details of microelectrode fabrication have been explained in [40]. The process started with deposition of 1.15 µm silicon oxide ($SiO_2$) by means of plasma-enhanced-chemical-vapor-deposition (PECVD) on top of the silicon substrate. The process followed by deposition of titanium/titanium nitrite (Ti/TiN) with thickness of 60 nm/30 nm. Finally, an aluminium/silicon/copper Al/Si/Cu (98/1/1 wt %) with thickness of 4.0 µm was deposited using physical vapour deposition (PVD). Photolithography was performed for transferring the square design on the Al/Si/Cu layer.

**Transfer of grown graphene to TAM**

The graphene grown using MFPLD and intercalated CVD graphene transferred through modified technique of graphene transfer using PMMA [41]. A solution of $FeCl_3$ was then used to etch the Ni and Cu catalyst, allowing the FLG and IFLG to be transferred to the desire position on the TAM surface. The PMMA sacrificial layer is eliminated through immersing graphene/PMMA in acetone vapor to minimize tearing of graphene which appears in the direct immersion in acetone solution, and brief acetone dipping for 2 min, followed by annealing at 500°C under gas mixtures of hydrogen and argon for 4 h to remove PMMA on graphene.

**Device characterization:**

Raman Spectroscopy measurements performed using a Raman spectrometer, Teksan (Takram P50C0R10, laser wavelength = 532 nm). Attenuated total reflection Fourier transform infrared (ATR-FTIR) spectra were collected with a Bruker Tensor 27 IR in room temperature deuterated triglycine sulphate (DTGS) detector, mid-IR source (4000 to 400 cm-1), and a KBr beam splitter. Each sample spectrum was collected for 15 scans with a resolution of 4 cm$^{-1}$ and a total acquisition time of 2.5 min. For the optical characterization of the device a Bruker IFS 66v/s FTIR and a calibrated blackbody source was used. For the responsivity measurement, FTIR test gave the relative response spectrum with respect to the detector. The set-up works the way that two detectors are collecting the same amount of irradiation from the infrared source, and then the absolute response was extracted from electrical signal when the actual sample is facing toward the known black body source with a narrow band filter. Background spectrum is measured by placing a DTGS pyroelectric detector at the same place as the measured photodetector.

## RESULTS AND DISCUSSION

The Raman spectroscopy of the graphene grown by MFPLD is shown in Figure 2. In table 1 the important peaks and ratios are shown. There were four peaks at around 1358, 1582, 2670, and 2890 cm$^{-1}$. The peak at 1358 cm$^{-1}$ is disorder peak (D). The G (graphitic) peak at ~1582 cm$^{−1}$ is associated with the doubly degenerate phonon mode ($E_2g$ symmetry) and is related to the in-plane C-C stretching in sp$^2$ sites inclusive of sp$^2$ chains and rings [42].

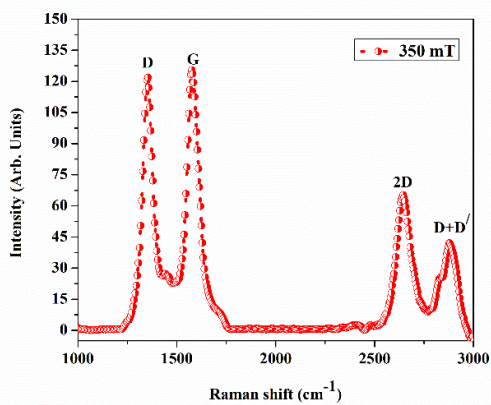

Figure 2 Raman spectroscopy of MFPLD grown graphene under 350 mT magnetic field.

The peak at around 2670 cm$^{-1}$ is the D-peak overtone and known as 2D peak. This peak is sensitive to number of layers and is closely linked to the details of the electronic band structure [40]. The $I_{2D}/I_G$ ratio and the 2D band shape show formation of FLG in MFPLD and it can be deduced that, there were less than 5 layers in all samples. The $I_{2D}/I_G$ ratio more than 50%, which corresponding to the growth of bi and tri layers graphene [43-45].

**Table 1.** Raman peaks intensity and position for $I_D$, $I_G$, $I_{2D}$, $I_{D+D'}$, $I_D/I_G$ and $I_{2D}/I_G$ of graphene grown by MFPLD under 350 mT magnetic field.

| Intensity / Growth condition | $I_D$(a.u) | $I_G$(a.u) | $I_{2D}$(a.u) | $I_{D+D'}$(a.u) | $I_D/I_G$ | $I_{2D}/I_G$ |
|---|---|---|---|---|---|---|
| **350 mT magnet** | 128 (1359) | 120 (1580) | 80(2645) | 44.5(2890) | 1.06 | 0.66 |

ATR-FTIR spectroscopy of the sample (not shown here), indicates that peak at around 1711 cm$^{-1}$ which belongs to C=O bond and normally is used to obtain the amount of oxygen in the graphene oxide is not presented in our sample. The lack of this peak in the grown structures indicates the lack of noticeable oxygen in the structure of the grown graphene.

Figure 3 gives Raman spectra of FLG/IFLG in three different zones. FeCl$_3$ has eight Raman-active modes 4A$_g$+4E$_g$ [46]. In FeCl$_3$ intercalant graphene, only four Raman modes, 2A$_{1g}$ and 2E$_g$, were observed at 93 cm$^{-1}$, 139 cm$^{-1}$, 181 cm$^{-1}$ and 287 cm$^{-1}$. The shift of the peaks as compared to bulk FeCl$_3$ agree with previous reports [39, 46] and further validate the intercalation process (Figure 3a). The position of G peak is used for the staging.

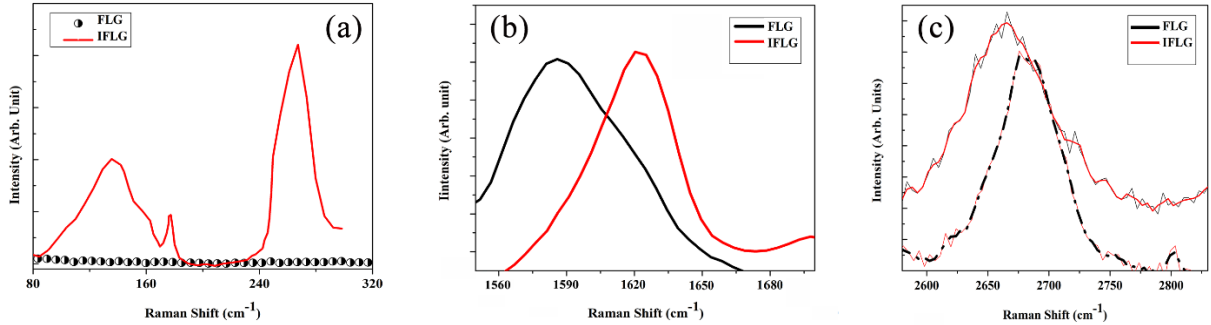

Figure 3. Raman spectra of FLG and IFLG measured for 532 nm excitation: (a) low frequency region; (b) d G-band region, and (c) 2D-band region.

The $I_{2D}/I_G$ ratio in the CVD graphene before intercalation is ~0.37, which is corresponding to the growth of 3–4 layers of graphene [47, 48]. G bands shift to ~1625 cm$^{-1}$ in IFLG in Figure 3b is also confirms the intercalation of for 3 or 4 layers of graphene. IFLG enhanced optical and electrical properties of graphene is a known phenomenon (graphexeter) [49]. This material can be used to define photo-active junctions with an unprecedented property. It is well known that the strong charge-transfer between graphene and FeCl$_3$ molecules induces large p-doping of graphene and drastically changes the carriers' dynamics. The FeCl$_3$ layer sandwiched between two graphene layers accepts electrons from both the layers, leading to hole-doped graphene [50, 51]. The Photodetector was fabricated through wet transfer of graphene grown by MFPLD and IFLG to the TAM surface. The scanning electron microscope (SEM) images of tapered resist profile etch using resist plasma etching process are shown in Figures 4.

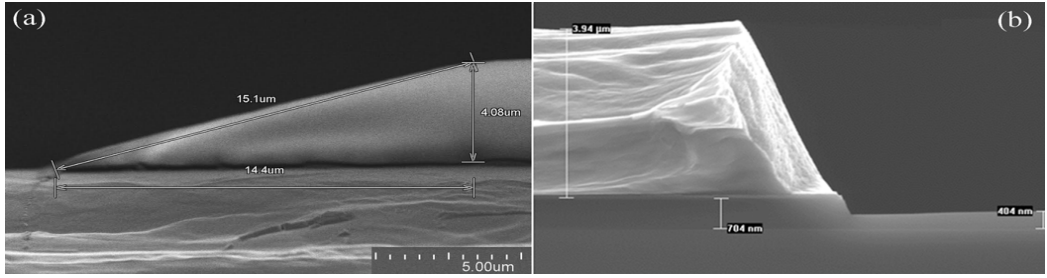

Figure 4. SEM image of (a) Tapered resist profile etch using resist plasma etching process (b) aluminium tapered microelectrode after aluminium etch process

Responsivity is a figure of merit of a photodetector and is defined as the output signal divided by the input optical power. If the output signal type is current, then the current responsivity is the ratio of photocurrent $I_{ph}$ and incident power:

$$R_{ph} = \frac{I_{ph}}{P_{in}} \quad \text{Eq (1)}$$

where $I_{ph}$ (Amper) is the output signal of the detector and $P_{in}$ (Watt) is the input optical power. A more useful parameter in measurement of the photodetector is quantum efficiency (QE) which describes the performance of by determining the percentage of photons that reach the detector, generate photo-carriers, and contribute to the photocurrent. QE is defined as the number of (e–h) pairs per second collected to produce the photocurrent $I_{ph}$, divided by the number of incident/absorbed photons per second. Therefore, it can be written as $QE = (I_{ph}/q)/\phi_{in}$ where q is the electron charge, $\phi_{in} = P_{in}/E_{ph}$. The incoming photon flux $\phi_{in}$ is related to incident photon energy $h\nu$ and power $P_{in}$, thus we get the following known equation (2) for QE calculation:

$$QE = \frac{h\nu}{q} R_{ph} \quad \text{Eq (2)}$$

The Responsivity and the QE of the photodetector spectra of the photodetector in NIR region at -0.5 negative bias voltages at 300 °K is presented in Figures 5(a) and (b) Respectively.

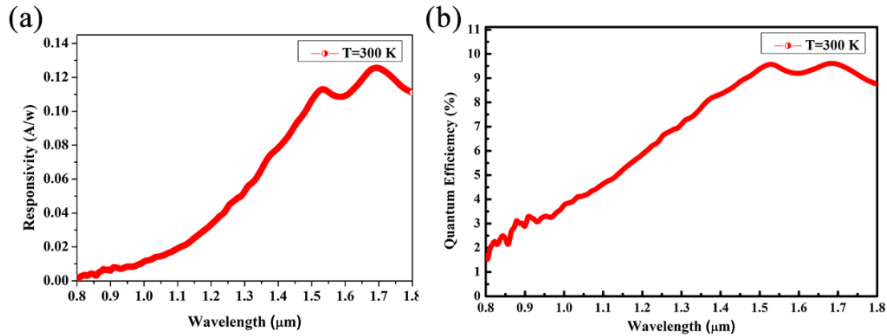

Figure 5. (a) Responsivity and QE spectra of the photodetector in NIR region at -0.5 negative bias voltages

The highest responsivity was obtained with a maximum of 130 mA/W at λ = 1700 nm and the corresponding 9.5% QE in the device under investigation. To find out the reason for the obtained results we need to find out the mechanism of photocurrent formation and transformation in IFLG and FLG. The intercalation of graphene with $FeCl_3$ modify the electronic transport properties in graphene through hole doped the graphene layers since it can accept electrons half from the top and half from the bottom graphene layers. In addition to this strong charge transfer doping, the intercalation increases the effective distance between the graphene layers and thus the electronic structure of IFLG becomes like that of SLG.

The additional electronic states introduced in IFLG facilitate photoexcited electron transition between lower and higher energy levels with the assistance of optical phonons. Optical phonons can be emitted and reabsorbed by the carriers in the higher energy states, thus slowing down the energy relaxation of the photoexcited electrons. This emission and reabsorption of optical phonons causes the number of photoexcited electrons to be built up slowly, resulting in a longer rise time [52]. The intercalants in the graphene structure act as a transient energy reservoir which results in the back-and-forth inelastic scattering of nonequilibrium electrons, causing a retardation of the internal thermalization. This effect already reported in the same structures that creates a strong photovoltage at the IFLG/FLG interfaces [49]. Photothermoelectric effect is the origin of the observed signal at the interface. At the interface there is an asymmetric heating and therefore, a temperature gradient is created. This leads to the diffusion of the hot carriers from the IFLG to the FLG layer. This charge motion yields a Seebeck voltage across the source-drain contacts, giving rise to the enhanced photothermoelectric effect inside each FLG. A record-high temperature difference of T = 5 $^0$K [53], is a reason that IFLG/FLG NIR photodetectors exhibit acceptable responsivity of R = 130 mV/W.

It should be noted that due to the significant shift in the Dirac point which is an indication of the metallic behavior of IFLG a very limited gate modulation effect is reported in the similar devices. However, applying a negative bias to the FLG contact enhances the induced injection of photo-excited holes from the IFLG into graphene. If similar metal contacts implemented for applying a bias voltage to the graphene photodetectors, because of the semi-metal behavior of graphene the bias voltage would drive a large dark current and a strong shot noise which is a serious limitation for the device performance.

With the application of an electric bias, the charge transfer takes place not only between $FeCl_3$ and graphene but also between Fe and Cl. Therefore, the $FeCl_3$ band can shift and contract as the magnitude of electric bias increases. Therefore, it is possible to control the hole doping of the $FeCl_3$ intercalation by the electric bias [35]. The tapered design of microelectrodes in TAM creates a nonuniform electric field through the interface of IFLG/FLG. This non uniform electric field is stronger in FLG/TAM and getting weaker as we go away from the contact. The nonuniform field sweep the created photo carriers toward the contact and improve the efficiency of photodetector. Additionally, this non uniform electric field catalyze the charge transfer between Fe and Cl by affecting on the dipole moments of Fe and Cl atoms and between $FeCl_3$ and FLG layer through the band gap opening of IFLG.

Specific detectivity (D*) was measured using the following equation [54, 55]:

$$D^* = R_i \left[2qJ_{dc} + \frac{4KT}{RA}\right]^{-1/2} \qquad (3)$$

where $J_{dc}$ is the dark current density and the RA is the differential resistance-area product. In this approach, the noise other than thermal and shot noise is ignored. To increase the D*, a balance and optimization for higher responsivity and corresponding quantum efficiency, lower dark current density, and larger RA are desired.

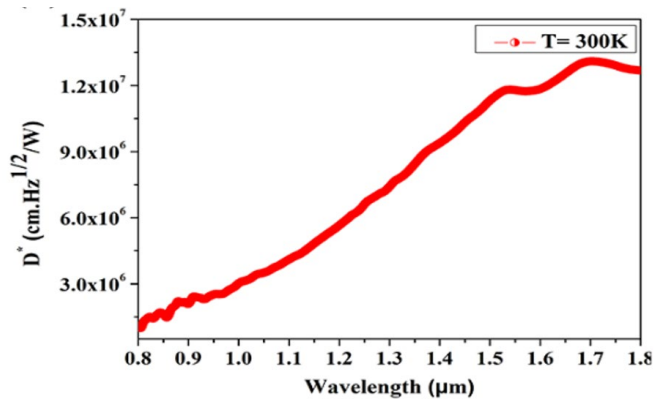

Figure 6. Specific detectivity, in NIR region at 300°K, vs wavelength

The specific detectivity of the device at 300°K is shown in Fig. 6. At 300 °K, the photodetector exhibits a maximum detectivity of 1.3 x 10$^7$ Jones at 1.7 μm NIR illumination. For the testing after processing, the samples were wire-bonded onto a leadless ceramic chip carrier (LCCC) and loaded into a liquid nitrogen cryostat for optical and electrical

characterizations(more information on measurement can be found in Ref[56]) . With the applying bias to the device and band gap opening of IFLG, it is expected that charge transfer between $FeCl_3$, graphene and IFLG/FLG increases since the IFLG resistivity is dropping with lowering the temperature.

In conclusion, we have successfully fabricated a full graphene-based photodetector for operation in the NIR region of the electromagnetic spectrum. The photoactive region of the detector was based on the FLG/IFLG junction formation. In order to improve the quality of base pristine graphene which acts as the main photogenerated carriers transport platform, we used a modify technique of MFPLD technique which improve the crystallinity of the grown graphene significantly. The TAM microelectrode also plays a critical role for band gap opening and fermi energy modification of IFLG. The photodetector presents an acceptable QE of around 10% and a responsivity of 130 mA/W at room temperature.

## REFERENCES

[1] S. Marconi *et al.*, "Photo thermal effect graphene detector featuring 105 Gbit s− 1 NRZ and 120 Gbit s− 1 PAM4 direct detection," *Nature communications,* vol. 12, no. 1, p. 806, 2021.
[2] M. G. Kwon *et al.*, "Performance enhancement of graphene/Ge near-infrared photodetector by modulating the doping level of graphene," *APL Photonics,* vol. 7, no. 2, 2022.
[3] A. Dehzangi, A. Haddadi, R. Chevallier, Y. Zhang, and M. Razeghi, "Fabrication of 12 μm pixel-pitch 1280 × 1024 extended short wavelength infrared focal plane array using heterojunction type-II superlattice-based photodetectors," *Semiconductor Science and Technology,* vol. 34, no. 3, p. 03LT01, 2019/02/04 2019, doi: 10.1088/1361-6641/aaf770.
[4] R. Chevallier, A. Dehzangi, A. Haddadii, and M. Razeghi, "Type-II superlattice-based extended short-wavelength infrared focal plane array with an AlAsSb/GaSb superlattice etch-stop layer to allow near-visible light detection," *Opt. Lett.,* vol. 42, no. 21, pp. 4299-4302, 2017/11/01 2017, doi: 10.1364/OL.42.004299.
[5] E. O. Polat *et al.*, "Flexible graphene photodetectors for wearable fitness monitoring," *Science advances,* vol. 5, no. 9, p. eaaw7846, 2019.
[6] M. Esfandiari *et al.*, "Recent and emerging applications of Graphene-based metamaterials in electromagnetics," *Materials & Design,* vol. 221, p. 110920, 2022.
[7] M. P. Hansen and D. S. Malchow, "Overview of SWIR detectors, cameras, and applications," in *SPIE Defense and Security Symposium*, 2008, vol. 6939: SPIE, p. 11.
[8] A. Rogalski, "HgCdTe infrared detector material: history, status and outlook," *Reports on Progress in Physics,* vol. 68, no. 10, p. 2267, 2005.
[9] A. Dehzangi *et al.*, "Impact of annealing time on copper doping concentration for mid-wave and long-wave mercury cadmium telluride," *Journal of Electronic Materials,* vol. 54, no. 5, 2025.
[10] M. A. Kinch, "HgCdTe: Recent trends in the ultimate IR semiconductor," *Journal of Electronic Materials,* vol. 39, no. 7, pp. 1043-1052, 2010.
[11] M. Henini and M. Razeghi, *Handbook of Infrared Detection Technologies*. Elsevier Science, 2002.
[12] M. Razeghi, A. Haddadi, A. Dehzangi, R. Chevallier, and T. Yang, "Recent advances in InAs/InAs1-xSbx/AlAs1-xSbx gap-engineered type-II superlattice-based photodetectors," in *Proc. SPIE 10177, Infrared Technology and Applications XLIII, 1017705* 2017, vol. 10177: SPIE, pp. 1017705-1017705-11. [Online]. Available: http://dx.doi.org/10.1117/12.2267044.
[13] Y. Arslan, F. Oguz, and C. Besikci, "Extended wavelength SWIR InGaAs focal plane array: Characteristics and limitations," *Infrared Physics & Technology,* vol. 70, pp. 134-137, 2015.
[14] M. Razeghi *et al.*, "High-performance short-wavelength infrared photodetectors based on type-II InAs/InAs1-xSbx/AlAs1-xSbx superlattices," in *Proc. SPIE 9819, Infrared Technology and Applications XLII*, 2016, vol. 9819: SPIE, pp. 98190A-98190A-7. [Online]. Available: http://dx.doi.org/10.1117/12.2228611.
[15] M. Razeghi, A. Dehzangi, and J. Li, "Multi-band SWIR-MWIR-LWIR Type-II superlattice based infrared photodetector," *Results in Optics,* vol. 2, p. 100054, 2021/01/01/ 2021, doi: https://doi.org/10.1016/j.rio.2021.100054.
[16] A. Dehzangi, D. Wu, R. McClintock, J. Li, A. Jaud, and M. Razeghi, "Demonstration of Planar Type-II Superlattice-Based Photodetectors Using Silicon Ion-Implantation," *Photonics,* vol. 7, no. 3, p. 68, 2020. [Online]. Available: https://www.mdpi.com/2304-6732/7/3/68.
[17] G. A. Sai-Halasz, R. Tsu, and L. Esaki, "A new semiconductor superlattice," *Applied Physics Letters,* vol. 30, no. 12, pp. 651-653, 1977, doi: 10.1063/1.89273.
[18] A. Dehzangi, "Multiband strain balanced superlattice material system for third generation infrared detectors," in *2024 IEEE Research and Applications of Photonics in Defense Conference (RAPID)*, 14-16 Aug. 2024 2024, pp. 01-02, doi: 10.1109/RAPID60772.2024.10647036.
[19] Y. Xiong, T. Chen, and W. Feng, "Broadband self-powered photodetector based on the large-area continuous WS0. 9Se1. 1 film," *Optical Materials,* vol. 145, p. 114412, 2023.
[20] L. Gao *et al.*, "Emerging applications of MXenes for photodetection: Recent advances and future challenges," *Materials Today,* vol. 61, pp. 169-190, 2022.

[21] T. Veeder *et al.*, "Accelerating discovery of tunable optical materials (ATOM)," in *Image Sensing Technologies: Materials, Devices, Systems, and Applications XI*, 2024 2024, vol. 13030: SPIE, p. 1303002, doi: 10.1117/12.3015512.
[22] F. Larki, Y. Abdi, P. Kameli, and H. Salamati, "An effort towards full graphene photodetectors," *Photonic Sensors,* pp. 1-37, 2022.
[23] A. Dehzangi and H. Mohseni, "Infrared focal plane arrays based on two-dimensional materials: possibilities and challenges," *Infrared Technology and Applications L,* vol. 13046, pp. 318-327, 2024, doi: 10.1117/12.3026776.
[24] S. Ansari, S. Bianconi, C.-M. Kang, and H. Mohseni, "From Material to Cameras: Low-Dimensional Photodetector Arrays on CMOS," *Small Methods,* vol. 8, no. 2, p. 2300595, 2024.
[25] K. I. Bolotin *et al.*, "Ultrahigh electron mobility in suspended graphene," *Solid state communications,* vol. 146, no. 9-10, pp. 351-355, 2008.
[26] T. Mueller, F. Xia, and P. Avouris, "Graphene photodetectors for high-speed optical communications," *Nature photonics,* vol. 4, no. 5, pp. 297-301, 2010.
[27] R. R. Nair *et al.*, "Fine structure constant defines visual transparency of graphene," *science,* vol. 320, no. 5881, pp. 1308-1308, 2008.
[28] X. Cao *et al.*, "Band gap opening of graphene by forming heterojunctions with the 2D carbonitrides nitrogenated holey graphene, g-C3N4, and g-CN: electric field effect," *The Journal of Physical Chemistry C,* vol. 120, no. 20, pp. 11299-11305, 2016.
[29] D. A. Mylnikov *et al.*, "Infrared photodetection in graphene-based heterostructures: bolometric and thermoelectric effects at the tunneling barrier," *npj 2D Materials and Applications,* vol. 8, no. 1, p. 34, 2024.
[30] D. K. Samarakoon and X.-Q. Wang, "Tunable band gap in hydrogenated bilayer graphene," *ACS nano,* vol. 4, no. 7, pp. 4126-4130, 2010.
[31] C. H. Lui, Z. Li, K. F. Mak, E. Cappelluti, and T. F. Heinz, "Observation of an electrically tunable band gap in trilayer graphene," *Nature Physics,* vol. 7, no. 12, pp. 944-947, 2011.
[32] S. Ullah *et al.*, "Advances and trends in chemically doped graphene," *Advanced Materials Interfaces,* vol. 7, no. 24, p. 2000999, 2020.
[33] M. Indika Senevirathna, D. K. Samarakoon, R. Gunasinghe, X.-Q. Wang, and M. D. Williams, "Bandgap opening of ferric chloride intercalated graphene by applying small electric field," *AIP Advances,* vol. 14, no. 12, 2024.
[34] W. Zhao, P. H. Tan, J. Liu, and A. C. Ferrari, "Intercalation of few-layer graphite flakes with FeCl3: Raman determination of Fermi level, layer by layer decoupling, and stability," *Journal of the American Chemical Society,* vol. 133, no. 15, pp. 5941-5946, 2011.
[35] J. Nathaniel and X.-Q. Wang, "Tunable electron and hole doping in FeCl3 intercalated graphene," *Applied Physics Letters,* vol. 100, no. 21, 2012.
[36] W. Liu, J. Kang, and K. Banerjee, "Characterization of FeCl 3 intercalation doped CVD few-layer graphene," *IEEE Electron Device Letters,* vol. 37, no. 9, pp. 1246-1249, 2016.
[37] F. Larki, P. Kameli, H. Nikmanesh, M. Jafari, and H. Salamati, "The influence of external magnetic field on the pulsed laser deposition growth of graphene on nickel substrate at room temperature," *Diamond and Related Materials,* vol. 93, pp. 233-240, 2019.
[38] X. Chen, L. Zhang, and S. Chen, "Large area CVD growth of graphene," *Synthetic Metals,* vol. 210, pp. 95-108, 2015.
[39] M. S. Dresselhaus and G. Dresselhaus, "Intercalation compounds of graphite," *Advances in Physics,* vol. 30, no. 2, pp. 139-326, 1981.
[40] M. R. Buyong, F. Larki, Y. Takamura, and B. Y. Majlis, "Tapered microelectrode array system for dielectrophoretically filtration: fabrication, characterization, and simulation study," *Journal of Micro/Nanolithography, MEMS, and MOEMS,* vol. 16, no. 4, pp. 044501-044501, 2017.
[41] X. Li *et al.*, "Transfer of large-area graphene films for high-performance transparent conductive electrodes," *Nano letters,* vol. 9, no. 12, pp. 4359-4363, 2009.
[42] L. Malard, M. Pimenta, G. Dresselhaus, and M. Dresselhaus, "Raman spectroscopy in graphene," *Physics Reports,* vol. 473, no. 5-6, pp. 51-87, 2009.
[43] U. Kalsoom, M. S. Rafique, S. Shahzadi, K. Fatima, and R. ShaheeN, "Bi-tri-and few-layer graphene growth by PLD technique using Ni as catalyst," *Materials Science-Poland,* vol. 35, no. 4, pp. 687-693, 2017.
[44] K. Wang, G. Tai, K. Wong, S. Lau, and W. Guo, "Ni induced few-layer graphene growth at low temperature by pulsed laser deposition," *AIP Advances,* vol. 1, no. 2, p. 022141, 2011.
[45] G. Pan *et al.*, "Transfer-free growth of graphene on SiO2 insulator substrate from sputtered carbon and nickel films," *Carbon,* vol. 65, pp. 349-358, 2013.
[46] N. Caswell and S. Solin, "Vibrational excitations of pure FeCl3 and graphite intercalated with ferric chloride," *Solid State Communications,* vol. 27, no. 10, pp. 961-967, 1978.
[47] I. Kumar and A. Khare, "Multi-and few-layer graphene on insulating substrate via pulsed laser deposition technique," *Applied surface science,* vol. 317, pp. 1004-1009, 2014.
[48] A. T. Koh, Y. M. Foong, and D. H. Chua, "Comparison of the mechanism of low defect few-layer graphene fabricated on different metals by pulsed laser deposition," *Diamond and Related Materials,* vol. 25, pp. 98-102, 2012.
[49] F. Withers, T. H. Bointon, M. F. Craciun, and S. Russo, "All-graphene photodetectors," *ACS nano,* vol. 7, no. 6, pp. 5052-5057, 2013.

[50] A. De Sanctis, J. D. Mehew, M. F. Craciun, and S. Russo, "Graphene-based light sensing: fabrication, characterisation, physical properties and performance," *Materials,* vol. 11, no. 9, p. 1762, 2018.

[51] D. Zhan *et al.*, "FeCl3-based few-layer graphene intercalation compounds: single linear dispersion electronic band structure and strong charge transfer doping," *Advanced Functional Materials,* vol. 20, no. 20, pp. 3504-3509, 2010.

[52] X. Zou *et al.*, "Ultrafast carrier dynamics in pristine and FeCl3-intercalated bilayer graphene," *Applied Physics Letters,* vol. 97, no. 14, 2010.

[53] M. W. Shabbir and M. N. Leuenberger, "Plasmonically enhanced tunable spectrally selective NIR and SWIR photodetector based on intercalation doped nanopatterned multilayer graphene," *arXiv preprint arXiv:2111.05982,* 2021.

[54] R. C. Jones, "On the relation between the speed of response and the detectivity of lead sulfide photoconductive cells," *JOSA,* vol. 43, no. 11, pp. 1008-1013, 1953.

[55] A. Dehzangi, J. Li, and M. Razeghi, "Band-structure-engineered high-gain LWIR photodetector based on a type-II superlattice," *Light: Science & Applications,* vol. 10, no. 1, p. 17, 2021/01/14 2021, doi: 10.1038/s41377-020-00453-x.

[56] A. M. Hoang, A. Dehzangi, S. Adhikary, and M. Razeghi, "High performance bias-selectable three-color Short-wave/Mid-wave/Long-wave Infrared Photodetectors based on Type-II InAs/GaSb/AlSb superlattices," *Scientific Reports,* Article vol. 6, p. 24144, 04/07/online 2016, doi: 10.1038/srep24144.